\documentclass[twocolumn,superscriptaddress,prl]{revtex4}
\usepackage{graphicx}
\usepackage{color}
\usepackage{epsfig,subfigure}
\usepackage{amsmath,amssymb,latexsym}
\usepackage{ascmac}

\def\U#1{{\rm #1}} 

\newcommand{\bra}[1]{\langle #1 |}
\newcommand{\ket}[1]{| #1 \rangle}

\newcommand{\expect}[1]{\left\langle #1 \right\rangle}

\begin{document}
\setlength{\abovedisplayskip}{6pt} 
\setlength{\belowdisplayskip}{6pt} 
\title{
Robustness of round-robin differential-phase-shift quantum-key-distribution protocol against source flaws
}
\author{Akihiro Mizutani$^*$}
\affiliation{Graduate School of Engineering Science, Osaka University,
Toyonaka, Osaka 560-8531, Japan}
\author{Nobuyuki Imoto}
\affiliation{Graduate School of Engineering Science, Osaka University,
Toyonaka, Osaka 560-8531, Japan}
\author{Kiyoshi Tamaki}
\affiliation{NTT Basic Research Laboratories, NTT Corporation, 
3-1, Morinosato-Wakamiya Atsugi-Shi, 243-0198, Japan\\
${}^*$mizutani@qi.mp.es.osaka-u.ac.jp}

\begin{abstract}
Recently, a new type of quantum key distribution, called the round-robin differential-phase-shift (RRDPS) protocol 
[Nature 509, 475 (2014)], was proposed, where the security can be guaranteed without monitoring any statistics. 
In this paper, we investigate source imperfections and side-channel attacks on the source of this protocol. 
We show that only three assumptions are needed for the security, and no detailed characterizations 
of the source or the side-channel attacks are needed. This high robustness is another striking advantage of the 
RRDPS protocol over other protocols.
\end{abstract}
\maketitle

Quantum key distribution (QKD) enables two distant parties (Alice and Bob) to generate a key, which is secret 
from any eavesdropper (Eve). 
Since the invention of the first QKD protocol, Bennett Brassard 1984 protocol~\cite{bb84}, 
there have been proposed many QKD protocols for both discrete variable protocols~\cite{e91,b92,six,sarg,dps,cow} 
and continuous variable protocols~\cite{cv1,cv2}. 
One of the most important tasks in the security proof is to derive an upper bound on the information leakage to Eve. 
Conventionally, it has been believed that the information leakage can be estimated by monitoring some statistics 
by Alice and Bob during the quantum communication part of the QKD protocol~
\cite{shor,sixlo,b92sec,three,sargtamaki,dps09,bbmkoashi,tomncom,dps12,cowprl,cowsec}. 
Recently, a new type of protocol, the round-robin differential-phase-shift (RRDPS) protocol~\cite{nature} was proposed 
and surprisingly, the information leakage of this protocol is estimated without any monitoring, 
but it depends only on the state prepared by Alice. 
This property leads to some practical advantages, such as the better tolerance on the bit error rate and the 
fast convergence in the finite key regime~\cite{nature}. 
This protocol has attracted intensive attentions from theoretical works~\cite{ma,takesue}, 
and proof-of-principle experiments have been demonstrated~\cite{rrdpsex1,takesue,rrdpsex2,rrdpsex3}. 

In practice, there are some issues to be addressed to guarantee the security of the RRDPS protocol when it is actually implemented. 
These issues arise because there is a gap between the properties of the actual devices used in QKD systems and the mathematical 
model that the security proofs assume, which is also the case for all QKD protocols. 
Therefore, to bridge this gap is crucial for the implementation security, 
and many works have been devoted in this direction~\cite{tomncom,tomanth,mdi,marcosncom,loss,lucamarini,wang,mdima}. 
In the case of the RRDPS protocol, all the security analyses including the original proof~\cite{nature} 
and the recent works~\cite{ma,takesue} have made ideal assumptions on Alice's light source 
(for instance, phase modulations are assumed to be perfect and any side-channel attacks are excluded). 
Therefore, to consider the security proof accommodating source flaws is indispensable toward a practical and secure 
implementation of the RRDPS protocol. 

In this paper, we extend the security proof of~\cite{nature} to accommodate the source flaws. 
Surprisingly, we found that the security can be guaranteed based only on the three assumptions on Alice's source. 
These assumptions are on the probability of emitting the vacuum state, on the probability that $L$ light pulses contain 
more than a particular number of photons, and on the independence among the sending states. 
Importantly, no assumptions on the phase modulation or detailed specifications of imperfections and 
side-channel attacks on the source are needed. 
Even with these imperfections and side-channels, 
we show that the RRDPS protocol can distribute the key over longer distances. 
These results show that the RRDPS protocol is highly robust against the source flaws, 
which is another striking advantage of this protocol over other protocols. 

Before explaining the security of the RRDPS protocol with the flawed sources, 
we summarize the assumptions we made on the devices. 
First, as for Alice's side, she employs blocks of 
$L$ light pulses, and applies phase modulation $\theta^{(k)}_{a_k}$ ($1\leq k \leq L$) 
to each of the pulses depending on a randomly chosen bit $a_k\in\{0,1\}$. 
The assumptions on Alice's sending states are summarized as follows. \\
\textbf{A1}.
For every light pulse, the probability of the vacuum emission for the bit value $0(1)$ is upper and lower bounded by 
$p_{\U{U},0(1)}(0)$ and $p_{\U{L},0(1)}(0)$, respectively 
(see Section 1 in the Supplementary material for the discussions on the estimation of these bounds for some experimental setups).\\
\textbf{A2}. The $L$ pulses contain in total at most $\nu_{\U{th}}$ photons except for the probability $e_{\U{src}}$.\\
\textbf{A3}.
There is no quantum and classical correlation among the sending states,
and the system that purifies each of the sending states is possessed by Alice.

We note that when more detailed characteristics of the source is available, we can relax the assumption {\bf A3} to accommodate any
classical correlations among the sending pulses.
In general, a classically correlated $L$-pulse state is written as
$\hat{\rho}=\int p(\tau_1,...,\tau_L)\bigotimes^L_{k=1}\hat{\rho}_{\tau_k}\U{d}\tau_1...\U{d}\tau_L$,
where $\tau_k~(1\leq k\leq L)$ denotes an internal parameter in the source to decide the $k^{\U{th}}$-pulse state.
Here, the probability distribution $p(\tau_1,...,\tau_L)$ can be arbitrary as long as it satisfies
$\int p(\tau_1,...,\tau_L)\U{d}\tau_1...\U{d}\tau_L=1$.
If Alice knows the upper and the lower bounds on the vacuum emission probability of the state ${\hat \rho}_{\tau_k}$ 
and the upper bound on the
probability that the photon number contained in $\bigotimes^L_{k=1}\hat{\rho}_{\tau_k}$ exceeds a certain threshold
for any realization of $\tau_k$ ($k=1,....,L$),
then we can prove the security even if  the pulses are classically correlated in any manner.
As an example of such a case, we discuss the security when each of the sending pulse is in a coherent state
for any $\tau_k$ and for any $k$ (see Section 4 in the Supplementary material for more detail).
In the following discussion, we assume that Alice has no knowledge about such detailed characterizations, and we 
prove the security based only on the assumptions \textbf{A1}, \textbf{A2} and \textbf{A3}.

We emphasize that we do not make any assumptions on phase modulations. 
Obviously, in order to generate a secret key, $\theta^{(k)}_0$ and $\theta^{(k)}_1$ need to be 
controlled such that the resulting bit error rate is low enough. 
However, for the security proof, this precise control over the phase modulation is not needed: our security proof 
holds not only when the actual value of phase modulations $\{\theta^{(k)}_0, \theta^{(k)}_1\}$ do not coincide with $\{0,\pi\}$,
but also when Alice has no knowledge about $\theta^{(k)}_0$ and $\theta^{(k)}_1$. 
The assumption {\bf A2} requires that 
$\U{Pr}\Big[\sum^L_{k=1}n_k>\nu_{\U{th}}\Big]\leq e_{\U{src}}$ 
must be satisfied, 
where $n_{k}$ denotes the number of photons included in the $k^{\U{th}}$ pulse, 
which would be obtained if we measured it, and 
the L.H.S represents the probability that the total photon number existing in the $L$ pulses exceeds $\nu_{\U{th}}$. 
We also emphasize that we do not make the single-mode assumption on the pulse, 
and the mode can depend on the bit value. 
This includes, for instance, the following cases: 
(1) the polarization of the pulse depends on the chosen bit value and (2) Eve performs a Trojan-horse-attack (THA)~\cite{gisinTHA},
where she injects a strong light pulse to Alice's source to obtain some information on the source from the 
back-reflected pulse.

From the assumption of the independence among the sending states described in {\bf A3},
the $k^{\U{th}}$ sending state is expressed as a partial trace over the system $\U{A_n}$ of the following state 
$\ket{\Psi_{a_k,k}}_{\U{A_n,B}}=\sum_{w}\sqrt{c^{(k)}_{w,a_k}}
\hat{U}^{(k)}_{a_k}\ket{w^{(k)}_{a_k}}_{\U{A_n}}\ket{\varphi^{(k)}_{w,a_k}}_{\U{B}}$,
where  
$c^{(k)}_{w,a_k}$ are non-negative real numbers satisfying $\sum_{w}c^{(k)}_{w,a_k}=1$, 
$\hat{U}^{(k)}_{a_k}$ is an arbitrary unitary operator on the system $\U{A_n}$, 
$\{\ket{w^{(k)}_{a_k}}_{\U{A_n}}\}_w$ are orthonormal bases of the ancilla system, 
and $\{\ket{\varphi^{(k)}_{w,a_k}}_{\U{B}}\}_w$ are orthonormal bases
to diagonalize the density operator of the sending state
\begin{align}
  \hat{\rho}^{(k)}_{a_k}
  :=\U{tr_{\U{A_n}}}\ket{\Psi_{a_k,k}}\bra{\Psi_{a_k,k}}_{\U{A_n,B}}. 
\end{align}
Note that the system $\U{A_n}$ that purifies each of the sending states is possessed by Alice.
Then, for each trial, Alice sends $\otimes^L_{k=1}\hat{\rho}^{(k)}_{a_k}$ to Bob over the quantum channel. 
Note that in the original protocol~\cite{nature}, Alice sends $\otimes^L_{k=1}e^{\U{i}\pi{}a_k\hat{n}_k}\ket{\Psi}$
in each trial, where $\hat{n}_k$ is the number operator for the $k^{\U{th}}$ pulse, and $\ket{\Psi}$ is the $L$-pulse state
(contains at most $\nu_{\U{th}}$ photons) before performing a perfect phase modulation ($e^{\U{i}\pi{}a_k\hat{n}_k}$). 

As for the assumptions on Bob's side, they are the same as those made in the original security proof~\cite{nature}, 
that is, 
Bob uses detectors that can discriminate among the vacuum, a single-photon, and multi photons, 
and Bob has a random number generator (RNG).
Using devices with these assumptions,
we describe Bob's actual procedures in what follows.
Note that Bob's actual procedures are the same as those in the original protocol~\cite{nature}. 
Bob first splits
$L$ incoming pulses into two trains of pulses, shifts backwards only one train by $r$ that is chosen 
randomly from $\{1,...,L-1\}$. 
Then Bob lets each of the first $L-r$ pulses in the shifted train interfere with each of the last $L-r$ 
pulses in the other train with a 50:50 beam splitter, 
and performs a photon measurement with the two detectors. 
Each of these detectors corresponds to the bit value of 0 and 1, respectively. 
Bob takes note of the bit value when he observes a single-photon in the original $L$ pulses in total, 
and otherwise he discards the data. 
Also, he records in which time slot he obtained the single-photon, 
and he announces this time slot and $r$ over the classical channel. 
From those information, Alice obtains a sifted key $a_{k_{\U{d}}}\oplus a_{k_{\U{d}}+r}$, where $k_{\U{d}}$ denotes the 
time slot of the single-photon detection. 
Bob repeats this process for many blocks containing $L$ pulses.  

Under the assumptions listed above, we prove the security of the RRDPS protocol with the source flaws. 
We note that, for simplicity of the analysis, we consider that the
number of blocks containing $L$ pulses sent is asymptotically large. In the proof, 
we construct a virtual protocol that cannot be distinguished from the actual protocol from Eve's viewpoint. 
In this virtual protocol, Alice first prepares her virtual qubit virA, ancilla qubits and system B in the following state 
\begin{align}
  \ket{\Phi}_{\U{virA,A_n,B}}=2^{-L/2}\bigotimes^L_{k=1}\sum_{a_k=0,1}
  \ket{a_k}_{\U{virA}}\ket{\Psi_{a_k,k}}_{\U{A_n,B}}, 
\label{virstate}
\end{align}
and sends only system B to Bob over the quantum channel. 
Here, 
this state is in the tensor product due to the assumption {\bf A3}, and we define $\{\ket{0}, \ket{1}\}$ as the $Z$ basis state. 

Next, we explain Bob's measurement procedures for the virtual protocol. 
As explained above, in the actual protocol, Bob performs an interference measurement on 
$i^{\U{th}}~(1\leq i\leq L)$ and $j^{\U{th}}~(1\leq j\leq L)$ pulses,
where a difference of $i$ and $j$ is randomly chosen by Bob's RNG 
({\it i.e.,} $|i-j|=r$). 
In the virtual protocol, however, Bob does not perform such an interference measurement, but performs the measurement to 
determine which pulse contains a single-photon among the incoming $L$ pulses. 
In this virtual
            measurement, the index $i$ is determined by the location of the single photon ($1\leq{}i\leq{}L$), and the other
            index $j$ is determined as
          \begin{align}
j=i+(-1)^br~(\U{mod}~L),
\label{randomj}
\end{align}
where $r$ is randomly chosen from the RNG, and $b$ is randomly chosen from 0 or 1 by Bob.
            After obtaining the pair $\{i,j\}$, he announces $\{i,j\}$ to Alice over the classical channel. 
Note that Eve has a perfect control over $i$ because she can freely 
choose which pulse contains a single-photon, 
but she cannot control $j$ at all because $j$ contains the randomness Bob locally chooses. 
  The reason why Bob can choose $j$ as Eq.~(\ref{randomj}) is that the probability distributions of obtaining $i$ and $j$
  if Bob postselects the successful detection event ({\it i.e.,} only a single-photon is detected from the $L$ pulses) 
are exactly the same for both actual and virtual protocols for any eavesdropping~\cite{nature}. 
This means that the classical information available to Eve is the same between the two protocols. 
Therefore, combined with the equivalence between the virtual and actual protocols in Alice's side, we are allowed to 
discuss the security based on the virtual protocol. 
In the virtual protocol, Alice keeps all the $L$ virtual qubits and the ancilla qubits 
when Bob obtains the successful detection. 

The quantity we use to measure the leaked information is the so-called phase error rate~\cite{koashiuncertain}, 
which is related with the smooth max-entropy~\cite{tomncom,renner}. 
With the phase error rate $e_{\U{ph}}$, the key rate per transmission of one pulse is expressed as~\cite{gllp} 
\begin{align}
  R=Q[1-f_{\U{EC}}h(e_{\U{b}})-h(e_{\U{ph}})]/L,
\end{align}
  where
$Q$ denotes the single-photon detection probability in Bob's measurement, 
$f_{\U{EC}}$ is an error correction efficiency, 
and $e_{\U{b}}$ denotes the bit error rate in the protocol and $h(x)=-x\log_2x-(1-x)\log_2(1-x)$ as the binary entropy 
function.
Here, $h(e_{\U{ph}})$ represents the fraction of bits to be shortened in the privacy amplification step.
            Once the sifted bits are shortened according to this fraction,
            the phase error information that Eve used to have 
becomes totally useless for her guessing the generated key.

Our goal below is to estimate the upper bound on the phase error rate. 
For the estimation, we need to define the phase error rate, but before we give its definition, it is convenient to 
rewrite Eq.~(\ref{virstate}) as 
$\ket{\Phi}_{\U{virA,A_n,B}}=2^{-L}
\bigotimes^L_{k=1}\Big[\sqrt{2+d_{k}}\ket{+}_{\U{virA}}\ket{\Phi^+_{k}}_{\U{A_n,B}}
+\sqrt{2-d_{k}}\ket{-}_{\U{virA}}\ket{\Phi^-_{k}}_{\U{A_n,B}}\Big],$ 
where we define 
$\ket{\pm}=(\ket{0}\pm\ket{1})/\sqrt{2}$ as the $X$ basis state,\\
$d_{k}=\sum_{w,w'}
\sqrt{c^{(k)}_{w,0}c^{(k)}_{w',1}}
\Big({}_{\U{A_n}}\expect{w'^{(k)}_1|
  \hat{U}^{(k)\dagger}_{1}\hat{U}^{(k)}_{0}|w^{(k)}_0}_{\U{A_n}}\\
{}_{\U{B}}\expect{\varphi^{(k)}_{w',1}|\varphi^{(k)}_{w,0}}_{\U{B}}+\U{C.C.}\Big)$ and
$\ket{\Phi^\pm_{k}}_{\U{A_n,B}}=(\ket{\Psi_{0,k}}_{\U{A_n,B}}
\pm\ket{\Psi_{1,k}}_{\U{A_n,B}})/\sqrt{2\pm d_{k}}$. 
Here, C.C. stands for complex conjugate. 
The key parameter in the security proof is the vacuum emission probability of $\ket{\Phi_{k}^\pm}_{\U{A_n,B}}$, 
which is defined by $p^{(k)}_{\pm}(0)$ and given by (see Section 2 in the Supplementary material for the detail)
\begin{align}
p^{(k)}_{\pm}(0)=
\frac{\Big(\sqrt{p^{(k)}_0(0)}\pm\sqrt{p^{(k)}_1(0)}\Big)^2}{2\pm d_{k}}. 
\label{p+-}
\end{align}
Here $p^{(k)}_{0}(0)$ and $p^{(k)}_{1}(0)$
denote the vacuum emission probabilities of $\ket{\Psi_{0,k}}_{\U{A_n,B}}$ and
$\ket{\Psi_{1,k}}_{\U{A_n,B}}$, respectively. 
In the virtual protocol, Alice's task is to guess the outcome of the 
$X$ basis measurement on the $j^{\U{th}}$ virtual qubit. 
The position of the $j^{\U{th}}$ virtual qubit is randomly chosen from the $L-1$ virtual qubits according to Eq.~(\ref{randomj}). 
Then, the phase error rate is defined as a fraction that Alice obtains the measurement 
outcome $-$ in her $X$ basis measurement on her $j^{\U{th}}$ virtual qubit that is randomly chosen from the $L-1$ virtual qubits. 
In the phase error rate estimation, 
we consider the worst case scenario that if the total photon number contained in the $L$ pulses exceeds $\nu_{\U{th}}$, 
Bob surely detects such an event as a successful detection, and 
Alice obtains the measurement outcome $-$ on her $j^{\U{th}}$ virtual qubit. 
By combining this worst case scenario and thanks to the randomness of $j$ from Eq.~(\ref{randomj}), the phase error rate is given by
\begin{eqnarray}
e_{\U{ph}}=e_{\U{src}}/Q+(1-e_{\U{src}}/Q)n^-/(L-1),
\label{defeph}
\end{eqnarray}
where $n^-$ denotes the number of the virtual qubits resulted in the measurement outcome of $-$ 
among the $L-1$ virtual qubits. 

In the following, we explain how to estimate the upper bound on $n^-$. 
Here, we use the fact that the statistics of the $X$ basis measurement on the system virA 
is not affected by any operations conducted on the system B. 
Therefore, in order to estimate the upper bound on $n^-$, we are allowed to perform the photon number measurement (PNM) 
on all the $L-1$ sending pulses in system B in Eq.~(\ref{virstate}). 
This PNM is an off-line measurement, and is not performed in either of the actual and virtual protocols. 
Let us denote by $n_u$ and $\U{M}_X^{(u)}\in\{+,-\}$ 
the outcome of the PNM of the $u^{\U{th}}$ $(1\leq u\leq L-1)$ sending pulse and 
the $X$ basis measurement outcome performed on Alice's $u^{\U{th}}$ virtual qubit, respectively. 
From these measurement results $n_1,...,n_{L-1}$, we estimate the upper bound on $n^-$. 
For the later convenience, we decompose $n^-$ into 
\begin{align}
n^-=n^-_{\U{nonvac}}+n^-_{\U{vac}},
\label{defeph2}
\end{align}
where $n^-_{\U{nonvac}}~(n^-_{\U{vac}})$ denotes the number of $u$ that satisfies $n_u>0~(n_u=0)$ and 
$\U{M}_X^{(u)}=-$. 

Now, we calculate the upper bound on Eq.~(\ref{defeph2}). 
First, we consider to upper bound $n^-_{\U{nonvac}}$. 
In so doing, we consider two worst case scenarios. 
The first worst case scenario is that if the $u^{\U{th}}$ sending state includes more than zero-photon ({\it i.e.,} $n_u>0$), 
we regard $\U{M}_X^{(u)}$ as $-$. 
The second one is that $\nu_{\U{th}}$ photons are distributed over the $L-1$ pulses 
such that the number of pulses that contain no photon is minimized. 
With these two worst case scenarios, $n^-_{\U{nonvac}}$ is upper bounded by 
\begin{align}
n^-_{\U{nonvac}}\leq n_{\U{nonvac}}\leq\nu_{\U{th}},
\label{Anu}
\end{align}
where $n_{\U{nonvac}}$ denotes the number of $u$ that satisfies $n_u>0$ among the $L-1$ pulses. 
In Eq.~(\ref{Anu}), the first and the second inequalities are due to the first and the second worst case scenarios, respectively. 

Next, we show the upper bound on $n^-_{\U{vac}}$, which is given by (see Section 3 in the Supplementary material for the detail)
\begin{align}
&n^-_{\U{vac}}=0
\label{resultnvac1}\\
&(\U{if}~p^{(k)}_0(0)=p^{(k)}_1(0)~\U{holds~for~all}~k~(1\leq k\leq L)),\nonumber
\end{align}
\begin{align}
&n^-_{\U{vac}}\leq 
\frac{L-1-\nu_{\U{th}}}{2}
\U{max}\Big\{\frac{(\sqrt{p_{\U{U},0}(0)}-\sqrt{p_{\U{L},1}(0)})^2}{p_{\U{U},0}(0)+p_{\U{L},1}(0)},\nonumber\\
&\frac{(\sqrt{p_{\U{L},0}(0)}-\sqrt{p_{\U{U},1}(0)})^2}{p_{\U{L},0}(0)+p_{\U{U},1}(0)}\Big\}+\U{max}_{N_{\U{vacd}}}\{N_{\U{vacd}}t\}\nonumber\\
&=:n^-_{\U{vac,U}}
\label{resultnvac2}\\
&(\U{if}~p^{(k)}_0(0)\neq p^{(k)}_1(0)~\U{for~some~or~every}~k)\nonumber, 
\end{align}
where $1\leq N_{\U{vacd}}\leq L-1-\nu_{\U{th}}$ and $0<t$. 
Note that $N_{\U{vacd}}$ and $t$ are related with a failure probability 
$\epsilon$ (see Eq.~(17) in the Supplementary material) of the Chernoff bound~\cite{chernoff}. 
Below, we explain the above results in more detail. 

(i) The first case is that $p^{(k)}_0(0)= p^{(k)}_1(0)$ is satisfied for all $k$ ($1\leq k\leq L$). 
From Eq.~(\ref{p+-}), we obtain $p^{(k)}_-(0)=0$, and hence $n^-_{\U{vac}}=0$. 
This means that if $n_u=0$, $\U{M}^{(u)}_{X}$ is $+$ and never be $-$. 
By combining the results in Eqs.~(\ref{Anu}) and (\ref{resultnvac1}), the phase error rate is upper bounded by 
\begin{align} 
e_{\U{ph}}\leq\min\Big\{\frac{e_{\U{src}}}{Q}+\Big(1-\frac{e_{\U{src}}}{Q}\Big)\frac{\nu_{\U{th}}}{L-1},~0.5\Big\}, 
\label{eph1}
\end{align}
Note that this upper bound is exactly the same as the one in the original security proof~\cite{nature}. 

(ii) The second case is that  $p^{(k)}_0(0)\neq p^{(k)}_1(0)$ occurs for some or every $k$. 
First, we give an example of how this situation arises. 
Suppose that Eve performs a THA where she injects a strong pulse to Alice's source to obtain some information 
on the source from the back-reflected light. 
To prevent this THA, Alice needs to suppress the intensity of the back-reflected light, which can be accomplished by 
installing some optical filters or optical isolators~\cite{lucamarini}. 
However, one cannot perfectly suppress the intensity, and moreover optical components, such as phase modulators, 
may have polarization dependence~\cite{feihu3state}, which leads to the situation of
$p_0^{(k)}(0)\neq p_1^{(k)}(0)$. 
This means that even if $n_u=0$, we cannot conclude $\U{M}^{(u)}_{X}=+$. 
Therefore, $n^-_{\U{vac}}$ results in a non-zero value, and $e_{\U{ph}}$ is increased compared with the one in the case~(i). 
In the case~(ii), by combining Eqs. (\ref{Anu}) and (\ref{resultnvac2}), the phase error rate can be obtained as 
\begin{align}
e_{\U{ph}}\leq\min\Bigl\{\frac{p_{\U{err}}}{Q}+\Big(1-\frac{p_{\U{err}}}{Q}\Big)\frac{\nu_{\U{th}}+n^-_{\U{vac,U}}}{L-1},~0.5\Bigl\}. 
\label{eph2}
\end{align}
Here, we define $p_{\U{err}}:=e_{\U{src}}+\epsilon-e_{\U{src}}\epsilon$.

We emphasize that the phase error rate given in Eqs.~(\ref{eph1}) and (\ref{eph2}) are derived only from the three 
assumptions: {\bf A1, A2} and {\bf A3}. 
This property is one of the striking features of the RRDPS protocol because other 
protocols usually need more detailed specifications of imperfections~\cite{wang,loss} 
and Eve's side-channel attacks on the source~\cite{lucamarini}. 
In particular, the practical QKD systems are threatened by the THA~\cite{vadim}, 
and the recent work quantitatively shows that the key generation rate 
of the BB84 protocol is compromised by this attack~\cite{lucamarini}. 
More specifically,~\cite{lucamarini} shows that when the mean photon number of the back-reflected light is 
$\mu_{\U{out}}=10^{-2}$, the achievable distance of the secure key generation is decreased down to only 10 km, 
while it is about 150 km without the THA. 
The reason for this drastic degradation is that the phase error rate is exponentially 
increasing with the distance~\cite{gllp}. 
In the RRDPS protocol, however, even if Eve performs the THA with $\mu_{\U{out}}=10^{-2}$, the increase of 
$\nu_{\U{th}}$ is only about $L\mu_{\U{out}}$. 
Therefore, the increase of $e_{\U{ph}}$ ({\it e.g.,} $L=100$) is about $L\mu_{\U{out}}/(L-1)\sim 1\%$ regardless of 
the distance, 
implying that only small amount of the additional privacy amplification is needed. 
This shows the robustness that the RRDPS has against the side-channel attacks on the source. 

\begin{figure}[t] 
 \begin{center}
 \includegraphics[width=7cm,clip]{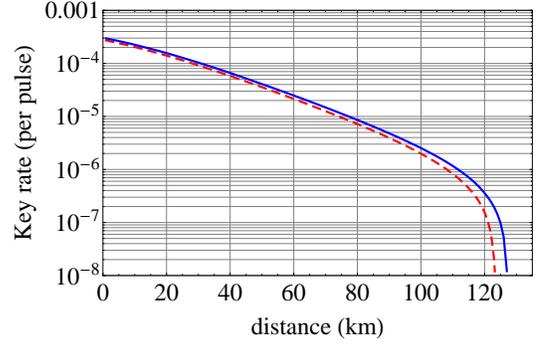}
\end{center}
\caption
{Secret key rate $R$ per pulse versus distances $l$. 
The solid line is for the case~(i):
$p^{(k)}_{0}(0)=p^{(k)}_1(0)$ is satisfied for all $k$, 
and the dashed line is for the case~(ii): $p^{(k)}_{0}(0)\neq p^{(k)}_1(0)$ occurs for some or every $k$.}
\label{fig2}
\end{figure}
Based on the above security proof, we show the key generation rate simulation results for the cases~(i) and (ii). 
In the simulation, we assume that Alice uses a weak coherent light source with the 
mean photon number $\mu_{0(1)}$ when she chooses the bit 0(1)~\cite{phase}. 
We set the channel transmittance as $\eta_{\U{ch}}=10^{-0.2l/10}$. 
In the detection side, 
we assume the detection efficiency and the dark count probability as $\eta_{\U{d}}=0.15$ 
and $p_{\U{d}}=5\times{}10^{-7}$, respectively. 
With these parameters, 
the successful detection probability that Bob detects the single-photon and the bit error rate are assumed to be given by 
$Q=(L\mu_0\eta_{\U{sy}})e^{-L\mu_0\eta_{\U{sy}}}/2+Lp_{\U{d}}$ and 
$e_{\U{bit}}=(L\mu_0\eta_{\U{sy}}e^{-L\mu_0\eta_{\U{sy}}}e_{\U{sym}}/2+Lp_{\U{d}}/2)/Q$, respectively. 
Here, $\eta_{\U{sy}}:=\eta_{\U{ch}}\eta_{\U{d}}$, and 
$e_{\U{sym}}$ is an overall misalignment error of the optical system, and we assume that $e_{\U{sym}}$ is 5\%. 
Also, we set $f_{\U{EC}}$ as 1.16. 

First, we show the simulation result for the case~(i). 
In this case, the mean photon number for both bits are the same $\mu_0=\mu_1=:\mu$. 
Under the above conditions, we plot the key rate $R$ with $L=128$ by the solid line in Fig. 1, 
where $R$ is optimized over the choice of $\mu$ and $\nu_{\U{th}}$ through the 
relation $e_{\U{src}}=1-\sum^{\nu_\U{th}}_{n=0}e^{-L\mu}(L\mu)^n/n!$. 

Next, we show the simulation result for the case~(ii). 
We assume that $\mu_1$ lies in the range $R_1:=[0.99\mu_0,1.01\mu_0]$. 
In this case, 
the upper and the lower bounds on $p^{(k)}_{0(1)}(0)$ are given by 
$p_{\U{U},0}(0)=e^{-\mu_0}$ ($p_{\U{U},1}(0)= e^{-0.99\mu_0}$) and $p_{\U{L},0}(0)=e^{-\mu_0}$ ($p_{\U{L},1}(0)=e^{-1.01\mu_0}$), respectively. 
Under these conditions, we plot the key rate $R$ with $L=128$ by the dashed line in Fig. 1, where $R$ is optimized over the choice of 
$\mu_0$, $\nu_{\U{th}}$ and $\epsilon$ through the relation 
$e_{\U{src}}=1-\min_{\gamma\in R_1}\{\sum^{\nu_{\U{th}}}_{n=0}e^{-L\gamma}(L\gamma)^n/n!\}$. 
This dashed line shows that even if $p^{(k)}_0(0)\neq p^{(k)}_1(0)$ occurs for some or every $k$, 
the degradation of the key generation rate is not so compromised. 
This result also shows the robustness of the RRDPS protocol against source flaws. 

To conclude, 
we have shown the security of the RRDPS protocol with imperfect light sources and side-channel attacks on Alice's source. 
In our security analysis, the characterization of Alice's source is simple in the sense that if Alice monitors 
only $\nu_{\U{th}}$, the vacuum emission probability
and the independence among the sending states, 
the amount of privacy amplification needed can be obtained. 
This means that the security of the RRDPS protocol can be guaranteed without detailed specifications of the source imperfections 
and side-channel attacks on the source. 
Moreover, we found that if the probabilities of emitting the vacuum state are the same for both bits, the phase error rate 
is exactly the same as the one in the original paper~\cite{nature}. 
Even if these probabilities differ, the performance of the key generation rate is not significantly compromised. 
These results show that the RRDPS protocol is highly robust against imperfections and side-channel attacks on the source, 
which is another practical advantage that this protocol has over other protocols. 

The authors thank H.-K. Lo, M. Koashi, H. Takesue, T. Sasaki, T. Yamamoto, K. Azuma, L. Qian, R. Ikuta, S. Kawakami
and G. Kato for fruitful discussions. 
AM and NI acknowledge support from the JSPS Grant-in-Aid for Scientific Research(A) 25247068. 
This work was in part funded by ImPACT Program of Council for Science, Technology and Innovation (Cabinet Office, Government of Japan).

\begin{widetext}
  \section
      {Supplemental Material}
      \maketitle
      \section{1.~Estimation of the vacuum emission probability}
\begin{figure}[t] 
 \begin{center}
 \includegraphics[width=11cm,clip]{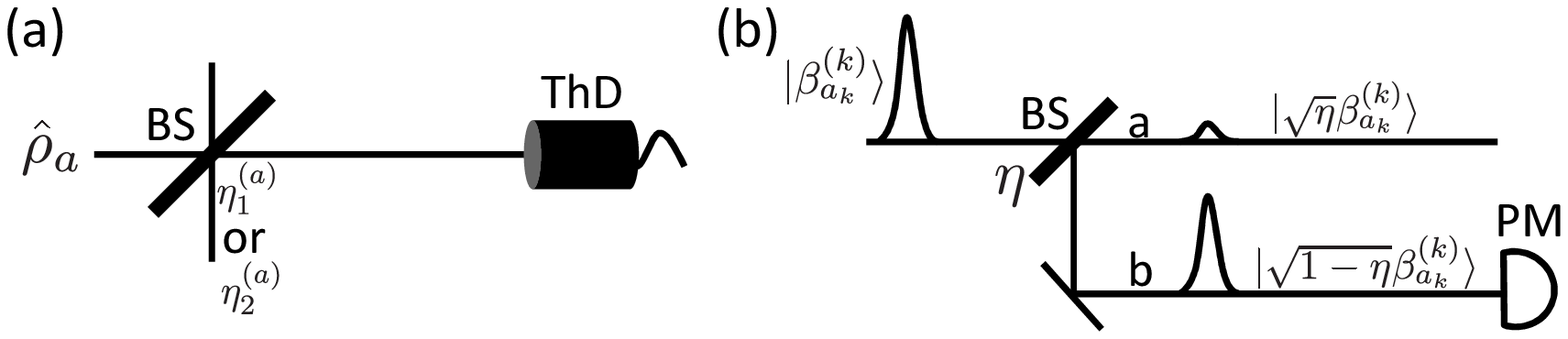}
\end{center}
\caption
{      (a)~Schematic of Alice's off-line measurement using 
      the detector decoy method~\cite{ddecoy}. This shows a detection setup that combines a variable beam splitter~(BS) of 
      transmittance $\eta^{(a)}_1$ or $\eta^{(a)}_2$~($\eta^{(a)}_1>\eta^{(a)}_2$) together with a threshold detector~(ThD).
      (b)~Schematic of Alice's on-line measurement when she uses a coherent light source. Alice firstly prepares a strong coherent
      light, after that she splits it by using a BS with transmittance $\eta~(\eta\ll1)$, and 
      monitors the intensity of the coherent light in mode b with a conventional power meter~(PM).
}
\end{figure}

As we have explained in the assumption A1 in the main text, Alice needs to estimate the upper and the lower bounds 
on the vacuum emission probability for each bit value $a=0,1$ ($p_{\U{L},a}(0)$ and $p_{\U{U},a}(0)$) in the actual experiments. 
Here, we propose how to estimate $p_{\U{L},a}(0)$ and $p_{\U{U},a}(0)$ for the following two particular cases. 
(i)
The first case is that Alice sends identical pulses when she chooses the same bit value. 
In this case, Alice performs an off-line measurement for the estimation. 
(ii) 
The second case is that Alice employs a coherent light source. 
In this case, we employ an on-line monitoring of the intensity of the sending light (see Fig.~1~(b)). 
Note that this method can be applied to the case
even if the photon number distribution is neither independent nor identical for each of the sending pulses. 
The following are the detailed explanations for (i) and (ii). 

(i)
First, we propose a method to estimate $p_{\U{L},a}(0)$ and $p_{\U{U},a}(0)$
under the case that Alice sends identical pulses if she chooses the same bit value {\it i.e.,}
$\hat{\rho}^{(s)}_{a_s}=\hat{\rho}^{(t)}_{a_t}$ is satisfied for any
$s,t\in\{1,...,L\}$ with $a_s=a_t$, where
$\hat{\rho}^{(k)}_{a_k}$ is defined in Eq.~(1) in the main text.
For later convenience, we denote $\hat{\rho}_a$ by the state when Alice chooses the bit value $a\in\{0,1\}$. 
Below, we explain how to estimate the bounds on the vacuum emission probability of $\hat{\rho}_a$ 
when Alice has a threshold detector. 
We consider to perform an off-line measurement to estimate the vacuum emission probability of $\hat{\rho}_a$
because the coherence between each photon number state of $\hat{\rho}_{a}$ is destroyed if Alice performs an on-line monitoring
of the photon number of each pulse.
In this measurement, if Alice has a threshold detector with unit efficiency and no dark count,
it is simple to obtain the bounds on the vacuum emission probability, because the number of the vacuum obtained in the measurement 
is the same as the number of the vacuum input to the threshold
detector. 
However, actual threshold detectors do not satisfy these conditions, and hence 
we use the {\it detector decoy method}~\cite{ddecoy} for the estimation of the vacuum emission probability of $\hat{\rho}_a$. 
In this method, 
Alice firstly places a beam splitter of transmittance $\eta^{(a)}_1$ before the threshold detector,
and counts the number of events where the vacuum outcome occurred (this number is denoted by $N^{(a)}_{1,\U{vac}}$) among
the number of emitted signals from Alice's source (this number is denoted by $N_1^{(a)}$). 
After obtaining $N^{(a)}_{1,\U{vac}}$, 
she repeats the same procedures with a beam splitter of transmittance $\eta^{(a)}_2$ ($\eta^{(a)}_1> \eta^{(a)}_2$), and
obtains $N^{(a)}_{2,\U{vac}}$ among $N_2^{(a)}$ (see Fig.1~(a)). 
With $N^{(a)}_{j,\U{vac}}$ for $j=1,2$, we have the following equation in the limit of asymptotically large $N_j^{(a)}$:
\begin{align}
  N^{(a)}_{j,\U{vac}}=N_j^{(a)}\U{tr}[\hat{\rho}_{a}\hat{\Pi}^{(a)}_{j,\U{vac}}],
  \label{ddm}
\end{align}
where $\hat{\Pi}^{(a)}_{j,\U{vac}}$ is a POVM element that corresponds to no click in the detector. 
Here, we assume that the POVM of the threshold detector with transmittance $\eta^{(a)}_{j}$ which contains two
elements $\hat{\Pi}^{(a)}_{j,\U{nonvac}}$ (this element corresponds to the click event in the detector)
, and $\hat{\Pi}^{(a)}_{j,\U{vac}}$
is given by $\hat{\Pi}^{(a)}_{j,\U{vac}}=(1-p_{\U{d}})\sum^{\infty}_{n=0}(1-\eta_j^{(a)}\eta_{\U{d}})^n\ket{n}\bra{n}$ and
$\hat{\Pi}^{(a)}_{j,\U{nonvac}}=\hat{I}-\hat{\Pi}^{(a)}_{j,\U{vac}}$, 
where $\eta_{\U{d}}$ and $p_{\U{d}}$ denote a detection efficiency and a dark count rate, respectively. 
In Eq.~(\ref{ddm}), $\hat{\rho}_{a}$ can be written as $\hat{\rho}_{a}=\sum^{\infty}_{n=0}p_{a}(n)\ket{n}\bra{n}$
with the Fock bases $\{\ket{n}\}$ without loss of generality,
because the off diagonal elements do not affect the measurement outcomes. 
By substituting the formulas of $\hat{\Pi}^{(a)}_{j,\U{vac}}$ and $\hat{\rho}_{a}$ into Eq.~(\ref{ddm}), $N^{(a)}_{j,\U{vac}}$
can be rewritten as 
\begin{align}
N^{(a)}_{j,\U{vac}}=N_j^{(a)}(1-p_{\U{d}})\sum^{\infty}_{n=0}p_{a}(n)(1-\eta_j^{(a)}\eta_{\U{d}})^n.
  \label{condition}
\end{align}
Note that if we replace $N^{(a)}_{j,\U{vac}}$ with $N^{(a)}_{j,\U{vac}}+\delta$ in Eq. (\ref{condition}) 
by using the Hoeffding inequality~\cite{hoeffding} or the Multiplicative Chernoff bound~\cite{marcosncom},
we can take into the finite-size effect $\delta$. 
By using Eq.~(\ref{condition}) for $j=1,2$,
we have that the upper and the lower bounds on the vacuum emission probability are respectively described as
\begin{align}
  p_{a}(0)&\geq \frac{(1-\eta_{\U{d}}\eta^{(a)}_1)N^{(a)}_{2,\U{vac}}/N^{(a)}_{2}-
    (1-\eta_{\U{d}}\eta^{(a)}_2)N^{(a)}_{1,\U{vac}}/N^{(a)}_1}{\eta_{\U{d}}(\eta^{(a)}_2-\eta^{(a)}_1)(1-p_{\U{d}})}, \\
  &=:p_{\U{L},a}(0),
  \label{pL0}
\end{align}
and
\begin{align}
  p_{a}(0)&\leq\min\Big\{\frac{N^{(a)}_{1,\U{vac}}}{(1-p_{\U{d}})N_1^{(a)}},\frac{N^{(a)}_{2,\U{vac}}}{(1-p_{\U{d}})N_2^{(a)}}\Big\}\\
  &=:p_{\U{U},a}(0).
    \label{pU0}
\end{align}
\newline

(ii)
Next, we consider how to estimate $p_{\U{L},a}(0)$ and $p_{\U{U},a}(0)$ for each bit value $a=0,1$ when Alice employs
a coherent light source {\it i.e.,} she knows the photon number distribution of each of the pulses as the Poissonian. 
Note that this method here can be applied to the case even if the photon number distribution
is not identical and independent for each of the sending pulses. 
In order to estimate the vacuum emission probability for the $k^{\U{th}}$ sending pulse ($1\leq k\leq L$), 
Alice firstly prepares a strong coherent light $\ket{\beta^{(k)}_{a_k}}$ with $|\beta|^2\gg1$, 
and she splits it by using a beam splitter with transmittance $\eta$ ($\eta\ll1$) as
$\ket{\beta^{(k)}_{a_k}}\rightarrow\ket{\sqrt{\eta}\beta^{(k)}_{a_k}}_{\U{a}}\ket{\sqrt{1-\eta}\beta^{(k)}_{a_k}}_{\U{b}}$.
The light in mode a is the $k^{\U{th}}$ sending light to Bob after the phase modulation, and the light in mode b
is an on-line monitoring light whose intensity is monitored by a conventional power meter. 
By this on-line monitoring, she obtains knowledge on the intensity of the coherent light before the beam splitter as 
$\beta^{-(k)}_{a_k}\leq\beta^{(k)}_{a_k}\leq\beta^{+(k)}_{a_k}$~(see Fig.1~(b)) for all $k$ ($1\leq{}k\leq~L$). 
Once Alice knows from the monitoring the range of $\beta^{(k)}_{a_k}$ for all $k$, 
we have that the bounds on the vacuum emission probability of the sending light in mode a are described as 
\begin{align}
  p_{\U{U},a}(0)=\max_{k\in\{1,...,L\}}e^{-\eta\beta^{-(k)}_{a_k}},\\
  p_{\U{L},a}(0)=\min_{k\in\{1,...,L\}}e^{-\eta\beta^{+(k)}_{a_k}}.
\end{align}
\newline

In the above estimations ((i) and (ii)), we have ignored any side-channels on the source {\it i.e.,} we explained the
estimation methods for $p_{\U{U},a}(0)$ and $p_{\U{L},a}(0)$ under the assumption that 
Alice's laboratory is perfectly protected from the environment outside. 
However, in the actual implementations, this assumption is not guaranteed because of Eve's side-channel attacks such as 
Trojan-Horse Attacks (THA)~\cite{gisinTHA}.
Therefore, in order to certify the assumption A1 in the practical implementations, 
we need to estimate $p_{\U{U},a}(0)$ and $p_{\U{L},a}(0)$ including the Trojan horse light,
and we explain this issue in what follows. 
In this attack, Eve injects a bright light pulse in the coherent state $\ket{\sqrt{\mu_{\U{in}}}}$
into Alice's apparatus through the optical fiber, and she obtains a back-reflected light containing the information
on the phase modulator. 
Recently, in order to prevent a particular THA, {\it passive architecture method} was proposed~\cite{lucamarini}. 
In this method, Alice places an optical isolator, an attenuator, and an optical filter
in order to suppress the intensity of the incoming light to her apparatus. 
Suppose that $\mu_{\U{out}}$  be the intensity of the back-reflected light from her apparatus, the upper and the lower bounds on the
vacuum emission probability (denoted by $p^{\U{THA}}_{\U{U},a}(0)$ and $p^{\U{THA}}_{\U{L},a}(0)$, respectively)
are given by the following modifications to $p_{\U{U}, a}(0)$ and $p_{\U{L}, a}(0)$
\begin{align}
  p^{\U{THA}}_{\U{U},a}(0)=e^{-\mu_{\U{out}}}p_{\U{U},a}(0),\\
  p^{\U{THA}}_{\U{L},a}(0)=e^{-\mu_{\U{out}}}p_{\U{L},a}(0),
\end{align}
respectively.
Therefore, the vacuum emission probability can be estimated even when
the particular Trojan horse light affects on the vacuum emission probability.

\section{2.~derivation of $p^{(k)}_{+}(0)$ and $p^{(k)}_{-}(0)$}
Here, we derive Eq.~(5) in the main text, which is the vacuum emission probability of $\ket{\Phi^\pm_{k}}_{\U{A_nB}}$. 
This is calculated as
\begin{align}
  p^{(k)}_{\pm}(0)&
  =\U{tr}[\ket{0}\bra{0}_{\U{B}}\ket{\Phi^\pm_{k}}\bra{\Phi^\pm_{k}}_{\U{A_n,B}}]\\
&=||{}_{\U{B}}\expect{0|\Phi^\pm_{k}}_{\U{A_n,B}}||^2\\
  &=\frac{1}{2\pm d_{k}}\Big|\Big|
  \sum_{w}\Big(\sqrt{c^{(k)}_{w,0}}{}_{\U{B}}\expect{0|\varphi^{(k)}_{w,0}}_{\U{B}}
\hat{U}^{(k)}_0\ket{w^{(k)}_0}_{\U{A_n}}
\pm{}
\sqrt{c^{(k)}_{w,1}}{}_{\U{B}}\expect{0|\varphi^{(k)}_{w,1}}_{\U{B}}
\hat{U}^{(k)}_{1}\ket{w^{(k)}_1}_{\U{A_n}}\Big)\Big|\Big|^2.
\label{p+-deriv}
\end{align}
If $\hat{U}^{(k)}_0$ is chosen such that 
$\sum_w\sqrt{c^{(k)}_{w,0}}{}_{\U{B}}\expect{0|\varphi^{(k)}_{w,0}}_{\U{B}}
\hat{U}^{(k)}_0\ket{w^{(k)}_0}_{\U{A_n}}$ 
becomes parallel to
$\sum_w\sqrt{c^{(k)}_{w,1}}{}_{\U{B}}\expect{0|\varphi^{(k)}_{w,1}}_{\U{B}}
\hat{U}^{(k)}_{1}\ket{w^{(k)}_1}_{\U{A_n}}$, 
and the inner product of these two states becomes positive, 
Eq.~(\ref{p+-deriv}) leads to 
$p^{(k)}_{\pm}(0)=\Big(\sqrt{p^{(k)}_0(0)}\pm\sqrt{p^{(k)}_1(0)}\Big)^2/(2\pm d_{k})$,
where we use
$p^{(k)}_{a_k}(0)=\U{tr}[\hat{\rho}^{(k)}_{a_k}\ket{0}\bra{0}_{\U{B}}]
=\Big|\sum_{w}\sqrt{c^{(k)}_{w,a_k}}{}_{\U{B}}\expect{0|\varphi^{(k)}_{w,a_k}}_{\U{B}}\Big|^2$. 

\section{3. estimation of $e_{\U{ph}}$ when $p^{(k)}_{0}(0)\neq p^{(k)}_{1}(0)$ occurs for some or every $k$}
Here, we derive Eq.~(10) in the main text, which is the upper bound on $n^-_{\U{vac}}$ when 
$p^{(k)}_0(0)\neq p^{(k)}_1(0)$ occurs for some or every $k$. 
In so doing, we construct a stochastic trial. 
For this, we introduce the random variable for the $l^{\U{th}}$ trial $(1\leq l\leq N_{\U{vacd}})$ as
\begin{align}
X^{(l)}&=\begin{cases}
    1 & (\U{if}~\U{M}_X^{(l)}=-) \\
    0 & (\U{if}~\U{M}_X^{(l)}=+).
  \end{cases}
\label{rv}
\end{align}
Here, $N_{\U{vacd}}$ corresponds to the number of those instances with 
$p^{(k)}_0(0)\neq p^{(k)}_1(0)$ among the $L-1$ pulses. 
As explained in the assumption A1, since Alice does not know the exact probabilities of emitting the vacuum state for both bits, 
$N_{\U{vacd}}$ cannot be obtained in the actual experiments. 
What Alice knows about $N_{\U{vacd}}$ is just a range, which $N_{\U{vacd}}$ lies in:
\begin{align}
1\leq N_{\U{vacd}}\leq L-1-\nu_{\U{th}}. 
\label{n_vac}
\end{align}
Now, to obtain the upper bound on $n^-_{\U{vac}}$, we use the Chernoff bound~\cite{chernoff}. 
From this inequality, 
\begin{align}
&\U{Pr}[n^-_{\U{vac}}-\sum^{N_{\U{vacd}}}_{l=1}p(X^{(l)}=1)>N_{\U{vacd}}t]\leq \epsilon
\label{chernoff}
\end{align}
is obtained for any $t\in [0,1-p_{\U{ave}}]$, where $p_{\U{ave}}:=1/N_{\U{vacd}}\sum^{N_{\U{vacd}}}_{l=1}p(X^{(l)}=1)$. 
Here, the parameter $\epsilon$ is given by 
\begin{align}
\epsilon=\exp[-D(p_{\U{ave}}+t||p_{\U{ave}})N_{\U{vacd}}], 
\label{epsilon}
\end{align}
where $D(p||q):=p\ln{}p/q+(1-p)\ln{}(1-p)/(1-q)$. 
\newline
Eq.~(\ref{chernoff}) means that 
\begin{align}
n^-_{\U{vac}}\leq\sum^{N_{\U{vacd}}}_{l=1}p(X^{(l)}=1)+N_{\U{vacd}}t
\label{resultchernoff}
\end{align}
holds with the probability $1-\epsilon$. 
Here, $p(X^{(l)}=1)$ represents the probability 
that Alice's $X$ basis measurement outcome on the $l$th virtual qubit is $-$ conditioned on the vacuum emission. 
This is calculated by using Eq.~(5) in the main text and Eq.~(\ref{rv}) as 
\begin{align}
p(X^{(l)}=1)&=p(\U{M}_X^{(l)}=-|n_l=0)\nonumber\\
&=\frac{p(\U{M}_X^{(l)}=-\wedge n_l=0)}{\sum_{S\in\{+,-\}}p(\U{M}_X^{(l)}=S\wedge n_l=0)}\nonumber\\
&=\frac{\Big(\sqrt{p^{(l)}_0(0)}-\sqrt{p^{(l)}_1(0)}\Big)^2}{2(p^{(l)}_0(0)+p^{(l)}_1(0))}.
\label{fraction}
\end{align}
Finally, by considering the upper bound on Eq.~(\ref{fraction}), 
we obtain the upper bound on Eq.~(\ref{resultchernoff}) as
\begin{align}
n^-_{\U{vac}}&\leq\frac{L-1-\nu_{\U{th}}}{2}
\U{max}\Big\{\frac{(\sqrt{p_{\U{U},0}(0)}-\sqrt{p_{\U{L},1}(0)})^2}{p_{\U{U},0}(0)+p_{\U{L},1}(0)},
\frac{(\sqrt{p_{\U{L},0}(0)}-\sqrt{p_{\U{U},1}(0)})^2}{p_{\U{L},0}(0)+p_{\U{U},1}(0)}\Big\}+\U{max}_{N_{\U{vacd}}}\{N_{\U{vacd}}t\}\nonumber\\
&=:n^-_{\U{vac,U}}, 
\label{vacresult}
\end{align}
where the maximization of $N_{\U{vacd}}$ is taken over the constraint in Eq.~(\ref{n_vac}).

\section{4.~Phase error rate estimation with coherent light source}
As we have explained on page 1 in the main text, 
when Alice has more detailed characteristics of the source, we can relax the assumption {\bf A3} to accommodate any
classical correlations among the sending states. 
In general, a classically correlated $L$-pulse state when Alice chooses an $L$-bit string as $a_1,...,a_L$ is expressed as 
\begin{align}
  \hat{\rho}^{(a_1,...,a_L)}_{\rm B}
  =\int p(\tau_1,...,\tau_L)\bigotimes^L_{k=1}\hat{\rho}^{(k)}_{a_k,\tau_k}\U{d}\tau_1...\U{d}\tau_L.
  \label{separable}
\end{align}
Here, $\tau_k$ ($1\leq k\leq L$) denotes the parameter to determine the $k^{\U{th}}$ sending state. 
$p(\tau_1,...,\tau_L)$ denotes a probability distribution to determine the realization of the $L$-pulse state
$\bigotimes^L_{k=1}\hat{\rho}^{(k)}_{a_k,\tau_k}$, and 
we do not need to characterize $p(\tau_1,...,\tau_L)$ except for the normalization condition 
$\int p(\tau_1,...,\tau_L)\U{d}\tau_1...\U{d}\tau_L=1$.
In order to accommodate the classical correlations, Alice needs to guarantee that the following two conditions are satisfied
for any realizations $\xi:=\tau_1,...,\tau_L$
($\xi$ denotes an internal parameter of Alice's source to determine the realization of the $L$-pulse state). 
The two conditions are
(i) the upper bound on the number of photons contained in the state $\bigotimes^L_{k=1}\hat{\rho}^{(k)}_{a_k,\tau_k}$
except for a probability $e_{\U{src}}$, 
and (ii) the upper and the lower bounds on the vacuum emission probability of each pulse ${\hat \rho}_{a_{k},\tau_{k}}^{(k)}$ 
for both bits $b=0,1$.

As an example of satisfying the conditions (i) and (ii),
here we consider the case where Alice knows that the sending $L$ pulses are coherent states for any $\xi$, 
and the mean photon number of all the sending states for Alice's choice of the bit $b\in\{0,1\}$ lies in the range
$[\mu^{(\U{min})}_b,\mu^{(\U{max})}_b]$. 
For deriving the phase error rate, we introduce a virtual protocol, which is the same as the actual protocol from Eve's viewpoint. 
In this protocol, before sending the $L$-pulse state to Bob, Alice randomly chooses a particular realization of the sending states
in the tensor product state according to the probability distribution $p(\xi):=p(\tau_1,...,\tau_L)$, 
and let $e^{(\xi)}_{\U{ph}}$ denote the phase error rate for the $L$-pulse state with the realization $\xi$. 
From the convexity structure of Eq.~(\ref{separable}), the phase error rate of $\hat{\rho}^{(a_1,...,a_L)}_{\rm B}$ can be written as
\begin{align}
  e_{\U{ph}}=\int p(\xi)e^{(\xi)}_{\U{ph}}\U{d}\xi,
  \label{coherentphase}
  \end{align}
and from Eq.~(\ref{coherentphase}), we trivially obtain the following inequality
\begin{align}
  e_{\U{ph}}\leq\max_{\xi} e^{(\xi)}_{\U{ph}}.
    \label{coherentphaseerror}
  \end{align}
Below, we first derive $e_{\U{ph}}^{(\xi)}$, and then we consider the upper bound on the R.H.S of Eq.~(\ref{coherentphaseerror}).
\newline
In order to derive $e_{\U{ph}}^{(\xi)}$, we use Eq.~(\ref{eph2}), and obtain the upper bound on $e^{(\xi)}_{\U{ph}}$ as
\begin{align}
  e^{(\xi)}_{\U{ph}}\leq
  \frac{p_{\U{err}}}{Q}+\Big(1-\frac{p_{\U{err}}}{Q}\Big)\frac{\nu^{(\xi)}_{\U{th}}+n^{-,{(\xi)}}_{\U{vac,U}}}{L-1},
  \label{ephxi}
\end{align}
where $\nu^{(\xi)}_{\U{th}}$ denotes the number of photons contained in the state $\bigotimes^L_{k=1}\rho^{(k)}_{a_k,\tau_k}$,
and from Eq.~(\ref{resultnvac2}), $n^{-,{(\xi)}}_{\U{vac,U}}$ in Eq.~(\ref{ephxi}) is given by
\begin{align}
n^{-,(\xi)}_{\U{vac,U}}=
\frac{L-1-\nu^{(\xi)}_{\U{th}}}{2}
\U{max}\Big\{\frac{\Big(\sqrt{p^{(\xi)}_{\U{U},0}(0)}-\sqrt{p^{(\xi)}_{\U{L},1}(0)}\Big)^2}{p^{(\xi)}_{\U{U},0}(0)+p^{(\xi)}_{\U{L},1}(0)}, 
\frac{\Big(\sqrt{p^{(\xi)}_{\U{L},0}(0)}-
  \sqrt{p^{(\xi)}_{\U{U},1}(0)}\Big)^2}{p^{(\xi)}_{\U{L},0}(0)+p^{(\xi)}_{\U{U},1}(0)}\Big\}+\U{max}_{N^{(\xi)}_{\U{vacd}}}\{N^{(\xi)}_{\U{vacd}}t\},
  \end{align}
where $1\leq N^{(\xi)}_{\U{vacd}}\leq L-1-\nu^{(\xi)}_{\U{th}}$ and $0<t$. 
$p^{(\xi)}_{\U{U(L)},b}(0)$ denotes the upper (lower) bound on the vacuum emission probability among the $L$-pulse state
for the bit value $b\in\{0,1\}$ conditioned that the internal parameter of Alice's source takes the value of $\xi$.
The upper bound on $e_{\U{ph}}^{(\xi)}$, which is also the upper bound of $e_{\U{ph}}$, can be readily obtained by taking
the upper bound on $\max_{\xi}\{\nu^{(\xi)}_{\U{th}}+n^{-,(\xi)}_{\U{vac,U}}\}$ as
\begin{align}
e_{\U{ph}}&\leq\frac{p_{\U{err}}}{Q}+\Big(1-\frac{p_{\U{err}}}{Q}\Big)\frac{\max_{\xi}(\nu^{(\xi)}_{\U{th}}+n^{-,(\xi)}_{\U{vac,U}})}{L-1}\nonumber\\
&\leq
\frac{p_{\U{err}}}{Q}+\Big(1-\frac{p_{\U{err}}}{Q}\Big)\frac{
\nu^{(\U{max})}_{\U{th}}+
\frac{L-1-\nu^{(\U{max})}_{\U{th}}}{2}
\U{max}\Big\{\frac{\Big(\sqrt{e^{-\mu^{(\U{min})}_0}}-\sqrt{e^{-\mu^{(\U{max})}_{1}}}\Big)^2}{e^{-\mu^{(\U{min})}_0}+e^{-\mu^{(\U{max})}_1}}, 
\frac{\Big(\sqrt{e^{-\mu^{(\U{max})}_0}}-\sqrt{e^{-\mu^{(\U{min})}_1}}\Big)^2}{e^{-\mu^{(\U{max})}_0}+e^{-\mu^{(\U{min})}_1}}\Big\}
+\U{max}_{N^{(\U{max})}_{\U{vacd}}}\{N^{(\U{max})}_{\U{vacd}}t\}}{L-1}.
  \end{align}
Here, $\nu^{(\U{max})}_{\U{th}}$ is the maximal number of photons contained in the $L$ pulses over all $\xi$, 
which is determined such that 
$e_{\U{src}}=1-\sum^{\nu^{(\U{max})}_{\U{th}}}_{\nu=0}e^{-\mu^{(\U{max})}}(\mu^{(\U{max})})^{\nu}/\nu!$ holds for $\mu^{(\U{max})}$ denoting
$\mu^{(\U{max})}:=\max_{b\in\{0,1\}}\mu_b^{(\U{max})}$, and the maximization of 
$N^{(\U{max})}_{\U{vacd}}$  is taken over the range $1\leq N^{(\U{max})}_{\U{vacd}}\leq L-1-\nu^{(\U{max})}_{\U{th}}$.

  \end{widetext}


\begin{thebibliography}{99}
\bibitem{bb84}
C. H. Bennett and G. Brassard, in 
{\it Proceedings of IEEE International Conference on Computers, Systems, and Signal Processing} (IEEE, New York, 1984), pp. 175–179. 
\bibitem{e91}
A. K. Ekert, Phys. Rev. Lett. {\bf 67}, 661 (1991). 
\bibitem{b92}
C. H. Bennett, Phys. Rev. Lett. {\bf 68}, 3121 (1992). 
\bibitem{six}
D. Bru\ss, Phys. Rev. Lett. {\bf 81}, 3018 (1998).
\bibitem{dps}
K. Inoue, E. Waks, and Y. Yamamoto, Phys. Rev. Lett. {\bf 89}, 037902 (2002). 
\bibitem{sarg}
V. Scarani {\it et al}, Phys. Rev. Lett. {\bf 92}, 057901 (2004). 
\bibitem{cow}
N. Gisin, G. Ribordy, H. Zbinden, D. Stucki, N. Brunner, and V. Scarani, arXiv:quant-ph/0411022.
\bibitem{cv1}
S. L. Braunstein and P. van Loock, Rev. Mod. Phys. {\bf 77}, 513 (2005).
\bibitem{cv2}
C. Weedbrook {\it et al}, Rev. Mod. Phys. {\bf 84}, 621 (2012). 
\bibitem{shor}
P. W. Shor and J. Preskill, Phys. Rev. Lett. {\bf 85}, 441 (2000).
\bibitem{sixlo}
H.-K. Lo, arXiv:quant-ph/0102138.
\bibitem{b92sec}
K. Tamaki, M. Koashi, and N. Imoto, Phys. Rev. Lett. {\bf 90}, 167904 (2003). 
\bibitem{three}
J.-C. Boileau {\it et al}, Phys. Rev. Lett. {\bf 94}, 040503 (2005).
\bibitem{sargtamaki}
K. Tamaki and H.-K. Lo, Phys. Rev. A {\bf 73}, 010302(R) (2006).
\bibitem{bbmkoashi}
M. Koashi {\it et al}, arXiv:0804.0891.
\bibitem{dps09}
K. Wen, K. Tamaki, and Y. Yamamoto, Phys. Rev. Lett. {\bf 103}, 170503 (2009).
\bibitem{tomncom}
M. Tomamichel {\it et al}, Nature Commun. {\bf 3}, 634 (2012).
\bibitem{dps12}
K. Tamaki, G. Kato, and M. Koashi, arXiv:1208.1995v1.
\bibitem{cowprl}
T. Moroder {\it et al}, Phys. Rev. Lett. {\bf 109}, 260501 (2012).
\bibitem{cowsec}
B. Korzh {\it et al}, Nature Photon. {\bf 9}, 163-168 (2015). 
\bibitem{nature}
T. Sasaki, Y. Yamamoto, and M. Koashi, Nature {\bf 509}, 475 (2014). 
\bibitem{ma}
Z. Zhang {\it et al}, arXiv:1505.02481v1. 
\bibitem{takesue}
H. Takesue {\it et al}, Nature Photon. {\bf 9}, 827-831 (2015).
\bibitem{rrdpsex1}
J.-Y. Guan {\it et al}, Phys. Rev. Lett. {\bf 114}, 180502 (2015). 
\bibitem{rrdpsex2}
  Y.-H. Li {\it et al}, arXiv:1505.08142.
\bibitem{rrdpsex3}
  S. Wang {\it et al}, Nature Photon. {\bf 9}, 832-836 (2015).
\bibitem{tomanth}
M. Tomamichel, and A. Leverrier, arXiv:1506.08458. 
\bibitem{wang}
X.-B. Wang {\it et al}, Phys. Rev. A {\bf 77}, 042311 (2008). 
\bibitem{mdi}
  H.-K. Lo, M. Curty, and B. Qi, Phys. Rev. Lett. {\bf 108}, 130503 (2012).
\bibitem{marcosncom}
M. Curty {\it et al}, Nature Commun. {\bf 5} 3732 (2014). 
\bibitem{loss}
K. Tamaki {\it et al}, Phys. Rev. A {\bf 90}, 052314 (2014).\\
A. Mizutani {\it et al}, New J. Phys. {\bf 17}, 093011 (2015).
\bibitem{lucamarini}
M. Lucamarini {\it et al}, Phys. Rev. X {\bf 5}, 031030 (2015). 
\bibitem{mdima}
Z.-Q. Yin {\it et al}, Phys. Rev. A {\bf 88}, 062322 (2013). 
\bibitem{gisinTHA}
N. Gisin {\it et al}, Phys. Rev. A {\bf 73}, 022320 (2006). 
\bibitem{koashiuncertain}
M. Koashi, arXiv:0505108. 
\bibitem{renner}
R. Renner, Security of Quantum Key Distribution, Ph.D. thesis, ETH Zurich (2005). 
\bibitem{gllp}
D. Gottesman {\it et al}, Quantum Inf. Comput. {\bf 4}, 325 (2004). 
\bibitem{chernoff}
H. Chernoff, Ann. Math. Sat. {\bf 23}, pp.493-507 (1952). 
\bibitem{feihu3state}
F. Xu {\it et al}, Phys. Rev. A {\bf 92}, 032305 (2015).
\bibitem{vadim}
N. Jain {\it et al}, New J. Phys. {\bf 16}, 123030 (2014). 
\bibitem{phase}
Note that the phase of the coherent light needs not be randomized.
\bibitem{ddecoy}
T. Moroder, M. Curty, and N. L$\ddot{\U{u}}$tkenhaus, New J. Phys. {\bf 11}, 045008 (2009). 
\bibitem{hoeffding}
W. Hoeffding, J. Amer. Statist. Assoc. {\bf 58} (301), 13-30 (1963).
\end{thebibliography}
\end{document}